# Hole emitter whispering galleries of photonic quantum ring

O'D. Kwon[1†], M. J. Kim[1], S.-J. An[1], D. K. Kim[1], S. E. Lee[1], J. Bae[1], J.H. Yoon[1]

[1]*Elec.Eng.Dept., Pohang University of Science & Technology, San 31 Hyojadong, Pohang 790-784, Korea.*

B. H. Park[2], J. Kim[2], J. Ahn[2], and S. Park[2]

[2]*Samsung Electronics Co., Nongseo-Ri, Ki-Heung, Yong-In 440-600, Korea.*

[†] *To whom correspondence should be addressed. E-mail:* odkwon@postech.edu

We report on the first observation of "hole" whispering gallery lasers from semiconductor microcavities with three dimensional optical confinement, with thresholds potentially reducible to micro-to-nano ampere regimes according to a quadratic size-dependent reduction, due to ideal quantum wire properties of the naturally formed photonic quantum rings before imminent recombination in a dynamic steady state fashion. If the device size grows over a critical diameter, the quantum ring whispering gallery then begins to disappear. However, cooperative small hole arrays like 256×256 quantum ring emitters avoid the criticality and open a possibility of constructing practical dense electro-pumped micro-to-nano watt emitter arrays, amenable to mega-to-giga ring emitter chip development via present fabrication techniques.

  Lord Rayleigh's "concave" whispering gallery (WG) mode phenomenon [1] has triggered the optoelectronic large-scale integration (LSI) circuit research for applications in next-generation photonic switching and interconnect technologies to remove the so called electronic bottleneck, by developing low-threshold two dimensional (2D) whispering gallery mode semiconductor lasers in the last decade [2-5]. The LSI quest for electro-pumped emitter arrays has also evolved into the array development of 1D vertical cavity surface emitting lasers, for instance the 16×16 array work with a limited performance due to thermal instabilities [6]. The laser array integrations in fact turned out to be inferior even to earlier decade-long efforts of the quantum well Stark shift modulator LSI developments undertaken since mid 1980s [7]. Recent reviews on the 3D optical interconnect technology now predict that the optoelectronic LSI dream may not be realized for another decade [8], while the current electronic LSI technology is beckoning to hit the ultimate quantum limit resolutions near Bohr radius. To overcome the above bottleneck issues with micro-to-nano ampere emitter capabilities, we propose and analyze 3D WG lasers of helix mode, activated by recombinant photonic quantum rings (PQRs) naturally generated in the "concave or convex" WG region of active quantum well planes without any artificial etching for the ring definition. We have earlier reported on the PQR microresonators of concave 3D WG modes made of cylindrical mesas, or traditional micropillars with smoothly etched sidewalls, fabricated mostly from wafers consisting of a few 80 Å (Al)GaAs quantum well active planes between p-type top and n-type bottom GaAs/AlAs Bragg reflectors for vertical optical confinement [9]. As shown previously these devices exhibit ultralow threshold and transparency currents reducible to the micro-to-nano ampere regime, and thermal stabilities at operating temperatures due to $\sqrt{T}$-dependent quantum wire spectral shift, which are crucial for mega-to-giga pixel chip operations, and also extendible to nanoemitter developments to enable the bottom-up nanosystem paradigm as discussed later. Regarding their nonuniform spectral distributions except the distinct multi-family behavior that happens in the device of diameter $\phi = 10$ $\mu$m, Fig. 1A represents a family of concave WG modes as a function of active diameter, whose systematic variation of intermode spacing is consistent with a well-defined angular quantization rule for the 3D toroidal cavity helix modes [10]. The PQR beam is radiating in 3D but surface-normal dominant fashion unlike the 2D WG lasers.

We now demonstrate that the hole WG (convex) modes are generated from microresonators of negative mesa (hole) type, fabricated from the same kind of wafers [11], which then behave like a bit fuzzy PQRs. Their spectra as shown in Fig. 1B exhibit weak multi-peaks but clamped to a mode equivalent to concave WG modes of $\phi \sim 40$ $\mu$m without any noticeable cavity-dependent variations, in marked contrast to the index-guided concave WG modes. We note that these hole-like PQRs are conceivable only for the vertical cavity semiconductor structures, where a carrier gain-guiding factor contributing to the 3D hyper WG confinement is more dominant. Although Fig. 1B apparently shows no significant confinements for most helical standing modes, we will shortly present a remarkable cooperative confinement effect in the case of dense PQR hole emitter arrays [12]. Fig. 1C now summarizes spectra for microhollow (concentric hole and mesa) devices of different diameters, where the systematic variation of intermode spacing is weak again for outer mesas as well as inner holes, while the multi-peak resonance effect is now enhanced as the device diameter decreases. For $\phi$ larger than 50 $\mu$m in general, in contrast to the concave WG modes deteriorating slightly (Fig. 1A), the inner hole WG resonance disappears drastically and the outer mesa WG resonance deteriorates significantly (Fig. 1C), indicating an intermodal competition between the two apparently distinctive hyper WG modes. The size related data altogether indicate a presence of critical diameter $\phi_c = 40 \sim 50$ $\mu$m associated with the gain-guiding effect of hyper WGs. The critical diameter $\phi_c$ will of course vary for different device structures and semiconductor materials other than AlGaAs.

The CCD image of an illuminant PQR laser ($\phi = 15$ $\mu$m) and a typical narrowed spectrum are shown in Fig. 2A. The circumferential index-guided optical confinement of the concave WG orbits, together with the two Bragg reflectors, makes the 3D

Rayleigh toroid with a lateral Rayleigh width defined by $W_{Rayleigh} = (\phi/2)(1 - n_{eff}/n)$, where n is the bulk refractive index and $n_{eff}$ is the effective azimuthal refractive index, along the perimeter as shown in the pie-shaped cross sectional schematic (Fig. 3, inset) [9]. Now the helical WG standing wave manifold, or more exactly double Laguerre-Gaussian helix modes, dynamically induces concentric PQRs for imminently recombinant carriers in the quantum well Rayleigh region. This in turn exhibits thresholds of the micro-ampere regime for $\phi < 10$ μm while the transparency currents entering the nano-ampere zone as marked in yellow (Fig. 3), and the $\sqrt{T}$-dependent thermal stabilities. It is attributed to a photonic (de Broglie) quantum corral effect, which imposes a $\lambda/2$ period concentric corral-like transient ordering upon the imminently recombinant quantum well carriers as elaborated further shortly (λ=emission wavelength).

For the convex WG rings, proton implantation (dosage: $10^{16}$ $cm^{-2}$) steps are often added (Fig. 2B) for device isolations in order to confine the majority of the carriers to be funnelled through the un-implanted annular regions. However, we now present a high density cooperative PQR hole array with no implantation, where the neighboring holes surrounding a given hole seem to function as an effective isolation boundary. For example, Fig. 2C is a picture of 256×256 hole ($\phi$=8 μm ; pitch=20 μm) arrays we have made so far for a mega-pixel laser chip development with the implantation steps skipped, where the superbly uniform emission pattern results with 0.4-ampere total threshold injection current. An estimate of $I_{th} \sim 6$ μA ($I_{tr} \sim 2$ μA) per hole suggests much better effective isolation in average than the implantation (Fig. 3), and it will be worth improving further by modern LSI fabrication techniques and optimal designs of hole pixel size ($\leq 1$ μm), appropriately designed periodicity, trench isolation, and so on.

In fact the above results strongly encourage us to create dense cooperative nanohole emitters with controlled periodicity extending from nanometer to submicron length scales for mega-to-giga chip developments, which may be a viable road to the practical nanotechnology in the future [13].

Fig. 3 now sums up size-dependent transparency and threshold currents for the hyper WG lasers of pillar- and hole-type PQRs. Obviously the nano-ampere zone is the territory for imaginary nanopillar and nanohole emitters, although they may not necessarily be lasers anymore. It also confirms that the present proton implantation method is not effective for the carrier confinement. In fact, the PQR holes fabricated inside mesas require far smaller thresholds which are then nearly indistinguishable from the regular PQR values. Overall, the quantum ring data are consistent with theoretical calculations: The transparency ($I_{tr}$:curve T) and threshold ($I_{th}$:curve A) current expressions for the case of PQRs occupying the annular Rayleigh region is given by $I_{th} = I_{tr} + I_i = N^{1D} \times W_{Rayleigh}/(\lambda/2n_{eff}) \times \pi\phi \times (e/\eta\tau) + I_i$, where $N^{1D}$ is the 1D transparency carrier density, τ the carrier lifetime, η the quantum efficiency, and $I_i$ stands for internal loss [9, 14]. The overall yellow zone confirms the unprecedented nano-ampere regime for the PQR nanoemitters, whose fabrications are to be routine and reproducible, although our data are a bit scattered as shown in Fig. 3 presently. We further add that in the case of $I_{tr}$=30~50 nA device ($\phi$=6 μm) prethreshold chaotic phenomena also appear [15]. For smaller $\phi$'s the active volume decreases below 0.1 $\mu m^3$, and with the cavity Q factor over 15,000 [16], the corresponding spontaneous emission coefficient $\beta$ will become appreciable enough for thresholdless lasing without a sharp turn-on threshold [2], which often occurs in the PQR light-current analyses. Regarding the implant-isolated holes, the wide-spread data suggest a fuzzy ring trend growing as the device shrinks due to the growing leaky implantation boundary effect, and the hole PQR threshold data are actually approaching the curve B, whose formula is derived for the mesa by assuming that the Rayleigh region is now nothing but a piece of annular quantum well plane of random recombinant carriers instead:
$$I = N^{2D} \times W_{Rayleigh} \times \pi\phi \times (e/\eta\tau)$$

Regarding the photonic (de Broglie) quantum corral effect imposing a λ/2 period transient ordering upon the imminently recombinant carriers, we recall a well-known electronic quantum corral image from room temperature scanning tunneling microscope studies of Au atomic island plane at a given bias, although the optical λ/2 period for GaAs semiconductor will be substantially larger than the electronic de Broglie spacing [17]. We note that the Rayleigh region of quantum well planes is deeply buried beneath a few micron thick AlAs/GaAs Bragg reflectors not accessible for the present surface tunneling analysis. However, recent experiments and modeling work on dynamic interactions between carriers and transient field in a quantum well plane is a close case in point, although the work rather addresses a spatiotemporal transverse migration of the transient light field instead [18]. It thus appears that the transient quantum wire-like features seem to persist within the relevant time scale through thermal fluctuations. For an ensemble of carriers randomly distributed in the regional quantum well plane of concentration $10^{12}$ $cm^{-2}$ for instance, tens-of-nm scale local field-driven drifts of given carriers to a neighboring imminent PQR site should generate the proposed PQR ordering for an imminent recombination event of annihilating electron-hole pairs (for example the two different Rayleigh rings schematically shown in Fig. 3). We expect the standing waves in the Rayleigh region to give rise to a weak potential barrier for such a dynamic electron-hole pair process, perhaps an opposite case of extremely shallow quantum well excitons at room temperature [19] where even the shallow barriers tend to assure at least one bound state according to square well quantum mechanics. Recent 3D polarization measurements of the PQR reveal that the polarization vectors of the surface normal components lie always tangential to the PQR perimeter indicating a stronger carrier-photon coupling when the dynamic carrier distributions are quantum- wire like [15].

On the other hand, bottom-up nanowire device investigations such as fluidically arranged molecular crossbars [20], and superlattice pattern transferred molecular crossbars [21] were reported recently to realize the next generation nanodevice circuit technologies. In contrast, the lack of adequate, reliable high-density nanoemitter array work, except a few isolated single nanoemitter studies [22], appears to hold further nanotechnical developments. Thus a certain top-down dense array development of emitters in nano-ampere nanometer regimes with controlled periodicity is needed to remove another bottleneck in implementing practical integrated nanosystems, esp. for optical interconnects in nanoelectronics.

In conclusion, we have described the hole PQR phenomena, indicating a great potential for creating a mega-to-giga-level interconnect technology as well as intermediate base interconnect nanotechnology for the bottom-up assembly of nanoscale electronic and optoelectronic devices. Based upon the additional helix WG behaviours like sharp 3D index-guiding we may also imagine an optoelectronic LSI circuitry where submicron scale ridges of helix WG waveguides with sharp corners [15, 23], and appropriate 3D coupling angles and links to coplanar PQR emitters and receivers, are realizable in practice if we employ a certain quantum well intermixing technology to the WG-converted waveguide links [24].

We thank R. E. Slusher, E. Yablonovitch, B. Widom, D. A. B. Miller and late I. Prigogine for their encouragements, and K. An, D. Ahn, and V. G. Minogin for discussions. We also thank Y. C. Kim, E. K. Lee, T. S. Jung and K. H. Kim for assistance. Supported by NRL, KOSEF, BK21 and Samsung Co. projects.

**Figure Legends**

**Fig. 1.** Spectra for WG devices of photonic quantum ring (PQR). (A) Concave WG showing size-dependent resonances away from critical diameter. (B) Convex WG modes without resonance enhancement. (C) Hollow WG modes: the smaller diameter devices begin to show enhanced resonance (black for outer mesas : red for inner holes).

**Fig. 2.** (A) Narrowed spectrum taken from a concave WG PQR mesa in normal direction at $I$=800 $\mu$A ($\phi$=10 $\mu$m with the single mode tendency for smaller $\phi$. (Inset) SEM micrograph of a small PQR mesa ($\phi$=3 $\mu$m) before polyimide planarization and electrode metallization steps; CCD image of an illuminant PQR. (B) SEM micrograph of a PQR hole; (a) etched shallow trench, (b) implant region, (c) un-implant region. (Inset) CCD image of an illuminant hole PQR at $I$=15 $\mu$A and the spectrum at $I$=1.5 mA whose linewidth involving a few submodes superimposed ($\phi$=7 $\mu$m). (C) Color-coded 256×256 PQR hole ($\phi$=8 $\mu$m) array emission at I=0.7 A (total); (Inset) a magnified portion (×600). (D) CCD image of the illuminent hole emitters (×600).

**Fig. 3.** Threshold curves A and B from PQR and quantum well formulae, respectively, with corresponding Rayleigh toroid schematics ($R_{in}$) defined by Rayleigh width) transparency curve T for the PQR case. Data for transparency (empty symbols) and threshold (solid symbols) currents: circles for PQRs (magenta) and triangles for hollow PQRs (cyan): squares for PQR holes (black). Data at 8 $\mu$m correspond to the case of 256×256 hole arrays (see the arrow).

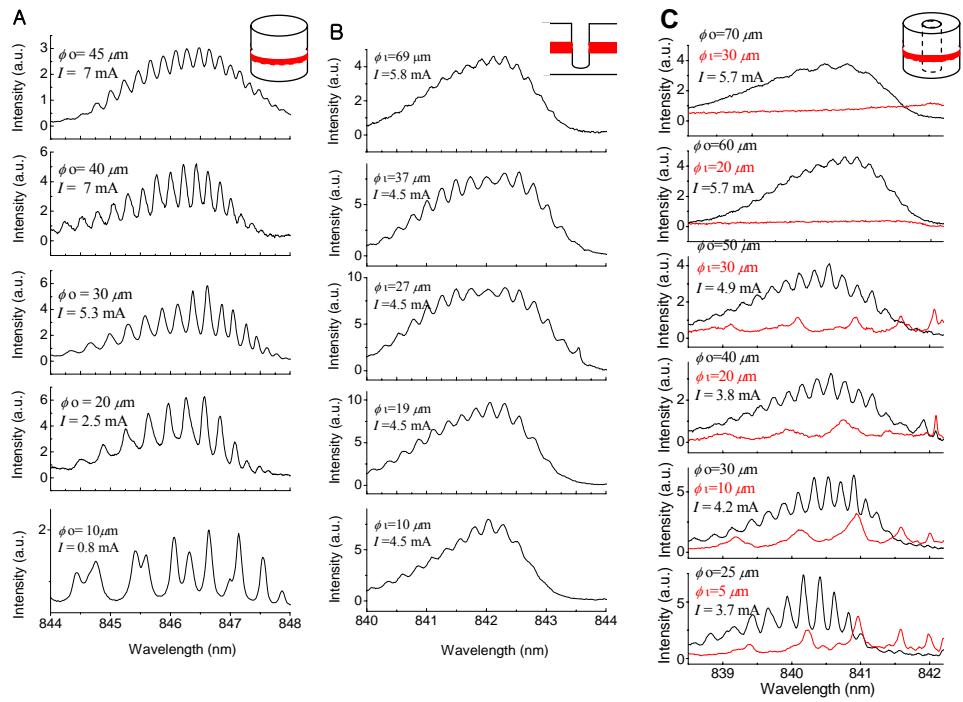

**Kwon_fig. 1**

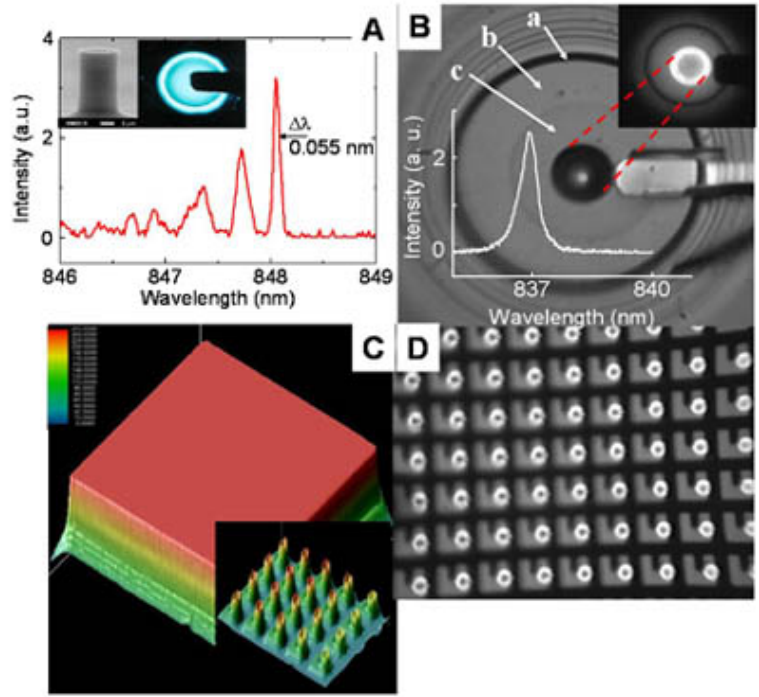

**Kwon_fig. 2**

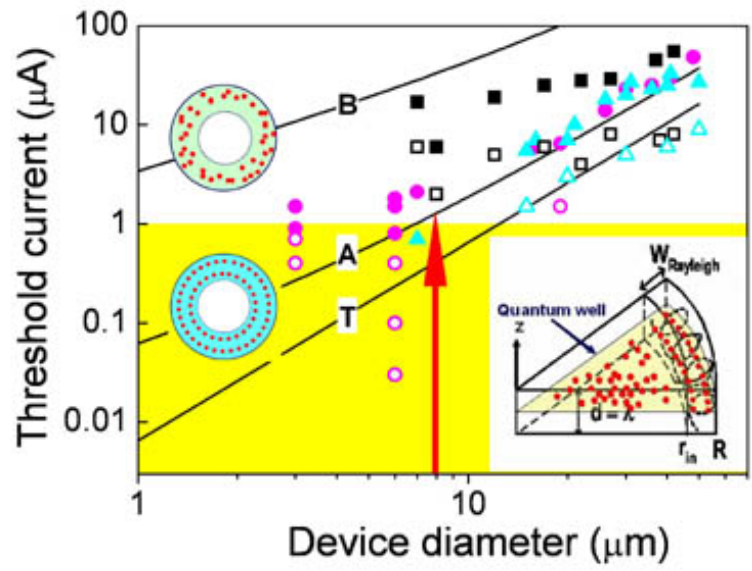

**Kwon_fig. 3**